# Attached Split-Ring Resonator Cavity for Magnon–Photon Coupling

Aram Akoi*, Liubov Ivzhenko, and Maciej Krawczyk+
Institute of Spintronics and Quantum Information, Faculty of Physics and Astronomy, Adam Mickiewicz University, Poznan, Poland

*aram.akoi@amu.edu.pl, +krawczyk@amu.edu.pl

## Abstract

We present a chip-scale planar cavity platform based on an attached split-ring resonator (ASRR) integrated with yttrium iron garnet (YIG) structures to achieve strong magnon–photon coupling in a compact hybrid system. The ASRR geometry was numerically optimized by tuning inter-ring spacing, gap width, substrate thickness, and permittivity, resulting in a quality factor of $Q \approx 190$ at 5.48 GHz, enabling strong microwave magnetic-field confinement and reduced radiative losses. The optimized cavity was coupled to YIG elements of three geometries: full ring, half ring, and disk. Full electromagnetic simulations show that the full-ring geometry exhibits balanced performance with coupling strength **115 MHz** and cooperativity **13.10**, while the half-ring shows a comparable coupling strength of **108 MHz** and slightly higher cooperativity **13.50**, despite edge-induced demagnetizing effects. In contrast, the disk geometry couples at lower bias magnetic fields and achieves the strongest interaction (**135 MHz, 25.30**), enabled by improved microwave magnetic-field overlap. These results demonstrate that geometry, rather than magnetic volume alone, is a key design parameter for tailoring magnon–photon coupling, providing a practical framework for lithography-compatible, on-chip hybrid magnonic and quantum devices.

## 1. Introduction

Cavity magnonics is an interdisciplinary field that bridges quantum optics, spintronics, and condensed matter physics. Its foundation is the coherent interaction between magnons, which are quantized collective spin-wave excitations in a ferromagnetic material, and microwave photons confined in a resonant cavity [1,2,3,4]. Magnon–photon coupling enables the reversible exchange of energy and information between spin and electromagnetic degrees of freedom, providing a versatile platform for applications such as quantum information processing, microwave-to-optical transduction, and non-reciprocal microwave devices [2,5,6]. Mode hybridization manifests as Rabi splitting in the transmission spectrum, giving rise to two hybrid modes known as magnon–photon polaritons. Realizing long-lived quantum memories and high-fidelity quantum transducers requires achieving strong coupling. In the strong-coupling regime, where the coupling strength $g$ exceeds the dissipation rates of the magnon $\kappa_m$ and photon $\kappa_c$ modes, coherent information transfer becomes possible because energy exchange between magnons and photons occurs faster than the typical decoherence processes [3–9].

Yttrium iron garnet (YIG), a ferrimagnetic insulator, is a preferred material for magnon–photon coupling due to its exceptionally low Gilbert damping ($\alpha \approx 10^{-5} - 10^{-3}$) and relatively high spin density ($\sim 4 \times 10^{27} m^{-3}$) [10-12]. Previous studies have shown that strong coupling can be achieved by using millimeter-scale YIG spheres in large 3D microwave cavities, making coherent coupling possible even at room temperature [3-7]. Nevertheless, the scalability of such geometries for realistic quantum and spintronic applications is limited because they are not practical for on-chip integration [1,2]. Planar resonators, such as split-ring resonators (SRRs), have emerged as a promising solution because of their compact form factor, high quality factors ($Q$), and ability to generate strong, localized microwave magnetic fields that can easily overlap with thin-film ferromagnets [13-17]. Because SRRs contain photons in a subwavelength resonant structure, they enhance mode overlap and strengthen the coupling between microwave photons and magnons, making them ideal for on-chip applications [13,14]. Most of the systems studied so far have considered SRR or inverse-SRR in a configuration detached from the feeding microwave line. However, an attached SRR could be more easily integrated and be useful for other applications, particularly in the field of magnonics, where omega-shape resonators are used for spin-wave detection [18-22] and have recently been used for the realization of the magnonic artificial neural network [23,24].

In this work, we propose and numerically optimize a planar SRR microwave cavity attached to the feeding line, named attached-SRR (ASRR), from the point of view of its $Q$, and compare it with two standard designs, i.e., separated-SRR (SSRR) and inverted-SRR (ISRR). The optimized ASRR is integrated with a ferromagnetic material (YIG) to calculate the magnon-photon coupling strength. We investigate how the geometry of the ferromagnetic element, in particular, a disk, a half ring, and a full ring, affects the magnon-photon coupling strength. We found that the ASRR is capable of supporting strong magnon-photon coupling in all considered YIG geometries, which makes the system suitable for further investigation towards application in hybrid magnonics and quantum technology.

## 2. Microwave Cavity Design

We consider three SRR configurations, ASRR, SSRR, and ISRR:

1. **Attached SRR (ASRR):** the feedline is directly connected to the outer ring at its gap (see Fig. 1a.1, and the transmission spectra in Fig. 1a.2).
2. **Inverted SRR (ISRR):** the feedline runs beneath the gap opening of the inner ring (see Fig. 1b.1 and 1b.2).
3. **Separated SRR (SSRR):** this configuration consists of concentric inner and outer rings with oppositely oriented gaps. The feedline is placed adjacent to the outer ring with a separation of 0.1 mm, producing inductive coupling (see Fig. 1c.1 and 1c.2).

All structures are constructed on a dielectric substrate with a thickness of 0.5 mm (we assume standard material RO4350B with permittivity $\varepsilon_r = 3.48$). The substrate is backed by a continuous copper ground plane on the bottom side, forming a microstrip configuration. The ASRR structures and the microstrip feedline are patterned on the top surface. The copper metallization is modeled with the conductivity of $\sigma_{cu} = 5.8 \times 10^7 \text{Sm}^{-1}$ (room-temperature value) with a perfect electric conductor (PEC) at the surfaces close to the dielectric substrate and normal boundary at the other surfaces of the Cu. This approximation is justified because, at microwave frequencies (~5.5 GHz), the skin depth of copper (~1 μm) is much smaller than the metal thickness (17 μm). However, this approximation neglects some ohmic losses and slightly overestimates the cavity quality factor. The microstrip feedline width is fixed at 1 mm to maintain a matched impedance, while the width of all inner SRR traces is kept constant at 0.5 mm. The separation between the inner and outer SRR rings is set to 0.2 mm, and each SRR gap width is maintained at 0.1 mm. To ensure that variations in resonant frequency and field distribution arise solely from differences in coupling configuration rather than changes in geometry or material properties, all parameters are kept identical across the comparative study of the three considered designs in the following part.

All simulations are performed using COMSOL Multiphysics (with RF Module), employing both the Frequency Domain and Eigenfrequency solvers to compute the electromagnetic response and linearized Landau-Lifshitz equations, respectively. In the frequency domain, Maxwell's equations are reduced to a vector wave equation for the electric field, assuming harmonic time dependence $e^{j\omega t}$. The governing equation solved by the RF module is:

$$\nabla \times \mu_r^{-1}(\nabla \times \boldsymbol{E}) - k_0^2\left(\epsilon_r - \frac{j\sigma}{\omega\epsilon_0}\right)\boldsymbol{E} = 0 \qquad (1)$$

where $\boldsymbol{E}$ is the electric field vector, $\mu_r$ is the relative permeability tensor, $\epsilon_r$ is the relative permittivity, $\sigma$ is the electrical conductivity, $\omega$ is the angular frequency, $\epsilon_0$ is the vacuum permittivity, and $k_0 = \omega/c$ is the free-space wavenumber. The magnetic field $\boldsymbol{H}$ is subsequently obtained from Maxwell–Faraday's law.

The microstrip feedline is excited using lumped ports with a characteristic impedance of 50 Ω. The outer boundaries of the computational domain are terminated using scattering boundary conditions, ensuring absorption of outgoing waves and preventing artificial reflections.

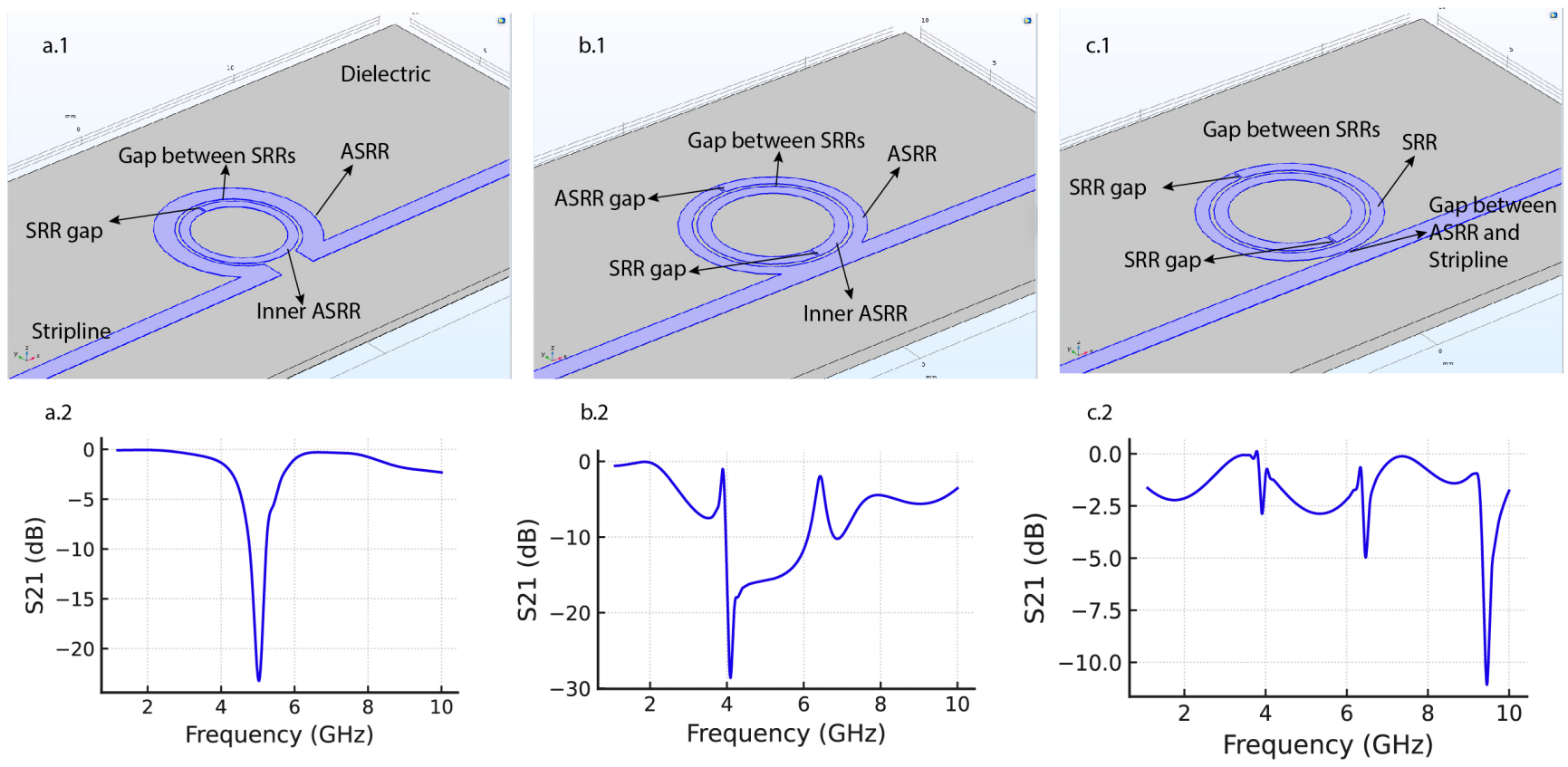

Figure 1: Considered SRRs and the respective $|S_{21}|$ spectra: (a) attached split-ring resonator (ASRR), (b) inverted SRR (ISRR), and (c) separated SRR (SSRR) geometries. The ASRR provides the highest quality factor ($Q$ = 109, 86, 45, for ASRR, ISRR, and SSRR, respectively).

The magnon–photon coupling is characterized by the coupling strength $g$ and the cooperativity $C$, defined as:

$$C = \frac{g^2}{\kappa_m \kappa_c}, \qquad (2)$$

where $\kappa_m$ and $\kappa_c$ are the magnon and photon decay rates, respectively. The strong-coupling regime is achieved when $C > 1$, ensuring that the coherent coupling rate dominates over dissipative losses in the system [3,17, 25]. The coupling strength $g$ depends on the number of spins in the ferromagnetic material and on the spatial overlap between the cavity microwave magnetic field and the magnetization dynamics. The cooperativity is inversely proportional to both decay rates. The magnon decay rate $\kappa_m$ was obtained from the full width at half maximum (FWHM) linewidth $\Delta f_m$ of the magnon resonance via $\kappa_m = 2\pi\Delta f_m$, consistent with standard cavity magnonics methods [4,6]. To avoid hybridization effects, the linewidth was evaluated in the uncoupled region of the dispersion, i.e., at low magnetic fields. The cavity decay rate $\kappa_c$ is determined from the FWHM of the uncoupled cavity resonance extracted from the transmission spectrum $|S_{21}|$, and is equivalently related to the quality factor as $\kappa_c = f_c/Q$, where $f_c$ is the cavity resonance frequency [6, 7, 26, 27]. Therefore, a high-Q cavity is essential, as it reduces cavity losses and increases the photon lifetime, thereby enhancing the effective interaction time between the microwave field and the magnon system. In this work, the cavity geometry is optimized to maximize the quality factor while maintaining the desired resonance frequency.

We compare the Q-factors of the three resonators by calculating their transmission spectra |S21| in the frequency range from 1 to 10 GHz, from which $Q$ is extracted. The spectra are shown in Fig. 1. Only the ASRR exhibits a single resonance within the considered frequency range, where |S21| reaches approximately −22 dB at the resonance frequency of 5.48 GHz. The quality factors of ASRR, ISRR, and SSRR are 109, 86, and 45, respectively. This demonstrates that the ASRR configuration is most suitable for magnon–photon coupling.

In the next step, we perform a systematic optimization of the ASRR parameters, in which individual geometrical and material parameters were varied independently while all other parameters were held constant to maximize the cavity quality factor. The results of the parametric analysis indicate that the Q-factor is most sensitive to the substrate thickness, inner-ring width, and inter-ring distance; each exhibits a distinct optimum and significant degradation away from that point. The split-gap width has a moderate influence (optimal near 0.10 mm), while varying the permittivity in the range 2.3–10 primarily shifts the resonance frequency and has a smaller effect on the Q-factor; the highest Q-factor is obtained for the chosen RO4350B substrate ($\varepsilon_r \approx 3.48$). Based on these parameter sweeps, the optimized geometry is selected as: inter-ring distance 0.275 mm, split gap 0.10 mm, inner-ring width 0.50 mm, and substrate thickness 0.50 mm, resulting in $Q \approx 190$ near 5.48 GHz. The transmission spectra and the magnetic field distribution at the resonance frequency of the optimized ASRR are shown in Fig. 2(a) and 2(b), respectively. The magnetic field is strongly confined between the rings and around the inner SRR region, on the side opposite to the split gap. This indicates that positioning a ferrimagnetic element inside the inner ring, close to the feedline, is favorable for achieving strong magnon–photon coupling.

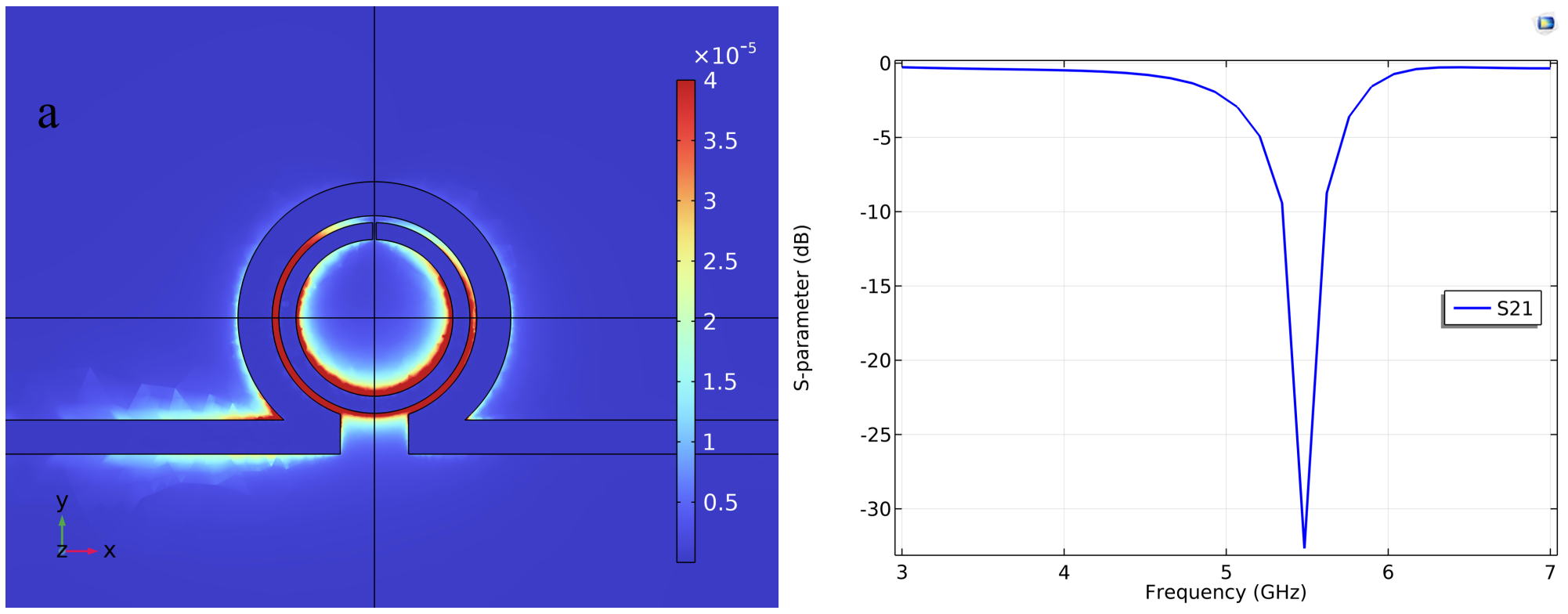

Figure 2: Optimized ASRR from the high Q factor point of view. *(a)* Spatial distribution of the microwave magnetic field (magnetic flux density) magnitude |B| at the cavity resonance frequency $f_{SRR}$ = 5.48 GHz. The maximum field value is approximately $6 \times 10^{-5}$ T located inside the inner SRR region, close to the feedline and opposite the split. (b) The transmission spectra with the single resonance at 5.48 GHz in the considered frequency range.

## 3. Ferromagnetic Resonance of YIG

In conventional ferromagnetic resonance (FMR) analysis, the Kittel formula is widely used to estimate the FMR frequency based on the applied static magnetic field $B_0$ and the saturation magnetization $M_s$ [28]. For the in-plane magnetized thin ferromagnetic film, it reads

$$f_m = \gamma/2\pi\sqrt{B_0(B_0 + \mu_0 M_s)} \qquad (3)$$

where $f_m$ is the resonance frequency, $\gamma$ is the gyromagnetic ratio, $\mu_0 M_s$ is the saturation magnetization, $\mu_0$ and is the vacuum permeability. $B_0 = \mu_0 H_0$ is the external magnetic field. However, the considered YIG samples, in the form of a disk, ring, and half ring, can exhibit non-uniform demagnetizing fields, particularly near the edges, where the geometry induces spatial variations of the internal magnetic field. To accurately determine the ferromagnetic resonance (FMR), we solve the Landau–Lifshitz–Gilbert (LLG) equation for the magnetization vector $\boldsymbol{M}$:

$$\frac{\partial \boldsymbol{M}}{\partial t} = -\gamma\mu_0 \boldsymbol{M} \times \boldsymbol{H}_{eff} + \frac{\alpha}{M_s}\boldsymbol{M} \times \frac{d\boldsymbol{M}}{\partial t}, \quad (4)$$

$\alpha$ is the Gilbert damping parameter, and $\boldsymbol{H}_{eff}$ is the effective magnetic field. The effective field, in our case, is given by

$$\boldsymbol{H}_{eff} = \boldsymbol{H}_0 + \boldsymbol{H}_{demag} + \boldsymbol{H}_{ex} \qquad (5)$$

where $H_0$ is the external magnetic field, we assume to be applied along the *x*-axis.

To obtain a tractable formulation, the LLG equation is linearized by expressing the magnetization as

$$\boldsymbol{M} = (M_s, m_y, m_z) \qquad (6)$$

where $|m_y|, |m_z| \ll M_s$ are small transverse dynamic components. Substituting this form into the LLG equation and neglecting higher-order terms in $m_y$ and $m_z$ yields a system of coupled linear equations describing the magnetization dynamics in the small-signal (linear response) regime. Assuming harmonic time dependence $m_y, m_z \sim e^{i\omega t}$, the problem reduces to an eigenvalue formulation for the angular frequency $\omega$. The demagnetizing field is computed using a scalar potential $\psi$, satisfying [29]:

$$\nabla^2 \psi = -\nabla . \mathbf{m} \qquad (7)$$

and then obtained as:

$$\boldsymbol{H}_{demag} = -\nabla \psi \, , \qquad (9)$$

while the exchange field is included as

$$\boldsymbol{H}_{ex} = \frac{2A_{ex}}{\mu_0 M_s} \nabla^2 \mathbf{m} \qquad (8)$$

where $A_{ex}$ is the exchange stiffness constant [30,31].

By sweeping the external magnetic field $H_0$, we determine the resonance magnetic field at frequency $f_m \approx f_c = 5.48$ GHz, ensuring spectral matching between magnon and photon modes. The YIG material parameters used in the simulations are as follows: saturation magnetization $M_s = 140$ kA/m, gyromagnetic $\frac{\gamma}{2\pi} \approx 28$ GHz/T, Gilbert damping constant $\alpha = 5 \times 10^{-4}$, and exchange stiffness $A_{ex} = 3.5 \times 10^{-12}$ J/m, consistent with typical values reported for high-quality YIG films [10–12]. For the YIG in the form of a ring (Fig. 3), and half-ring, the resonant bias field is similar and around 186 mT. For an in-plane magnetized YIG disk, we obtained $B_0^{disk} = 127$ mT.

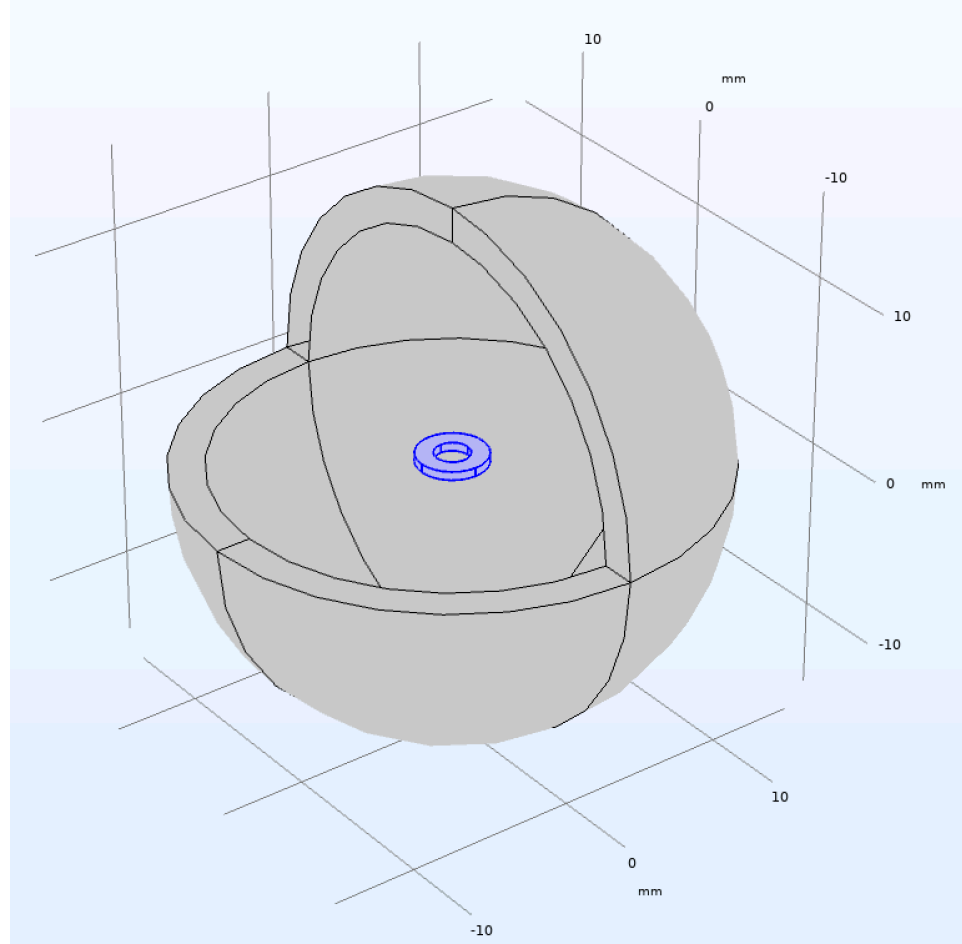

Figure 3: Geometry of the YIG ring sample in the simulation domain to illustrate the sample's shape and placement, and edges of the computational domain

## 4. Magnon-Photon Coupling Analysis

The strength of magnon-photon coupling of the ASRR loaded with elements of different shapes was examined using COMSOL Multiphysics with RF module, solving the electromagnetic wave equation, Eq. (1), with standard boundary conditions on the YIG surface. The Polder permeability tensor was used to introduce the description of magnetic response in the coupled ASRR-YIG simulation. It describes the gyromagnetic dynamics of the magnetization under a static bias magnetic field $B_0$ along the $x$-axis [30] it takes the form:

$$\hat{\mu}_r = \begin{pmatrix} 1 & 0 & 0 \\ 0 & \mu & -i\kappa \\ 0 & i\kappa & \mu \end{pmatrix}, \qquad (11)$$

where

$$\mu = 1 + \frac{(\omega_0 - i\alpha\omega)\omega_m}{(\omega_0 - i\alpha\omega)^2 - \omega^2}, \quad \kappa = \frac{\omega\omega_m}{(\omega_0 - i\alpha\omega)^2 - \omega^2}. \qquad (12)$$

Here, $\omega_0 = \gamma B_0$ (Larmor frequency), and $\omega_m = \gamma\mu_0 M_s$. The off-diagonal term κ represents the gyrotropic response arising from the precessional dynamics of magnetization, leading to non-reciprocal microwave propagation and Faraday rotation in microwave devices [30,31]. The Polder permeability tensor represents the linear dynamic susceptibility derived from the linearized LLG equation under the assumption of spatially uniform precession (long-wavelength limit).
Using the Polder tensor description for the coupling of magnetization dynamics with the microwave field is justified by the large YIG samples considered. While the Polder tensor assumes uniform precession, the standard electromagnetic boundary conditions introduced at the ferreomagnetic surfaces introduce nonuniform magnetization precession, representing magnetostatic spin waves [29, 32]. All tensor components were defined in the global coordinate system, and the complete anisotropic $\mu -$ tensor was manually entered in the material settings. The RF eigenfrequency solver, which solves wave equation with anisotropic permeability, was then used to obtain the transmission spectra of this hybrid structure. Strong spatial overlaps between the dynamic magnetization and the cavity microwave magnetic field and frequency matching (magnon and photon) are necessary for efficient coupling. Using the optimized ASRR, we positioned the YIG in three considered geometries inside the cavity, as shown in Fig. 4, where the strong localized magnetic field is present. Simulations were performed with the in-plane ($x$-axis) bias magnetic field $B_0$, which tunes the YIG FMR into resonance with the cavity, i.e., in the range of 165 – 195 mT for ring/half-ring and 110 – 150 mT for the disk.

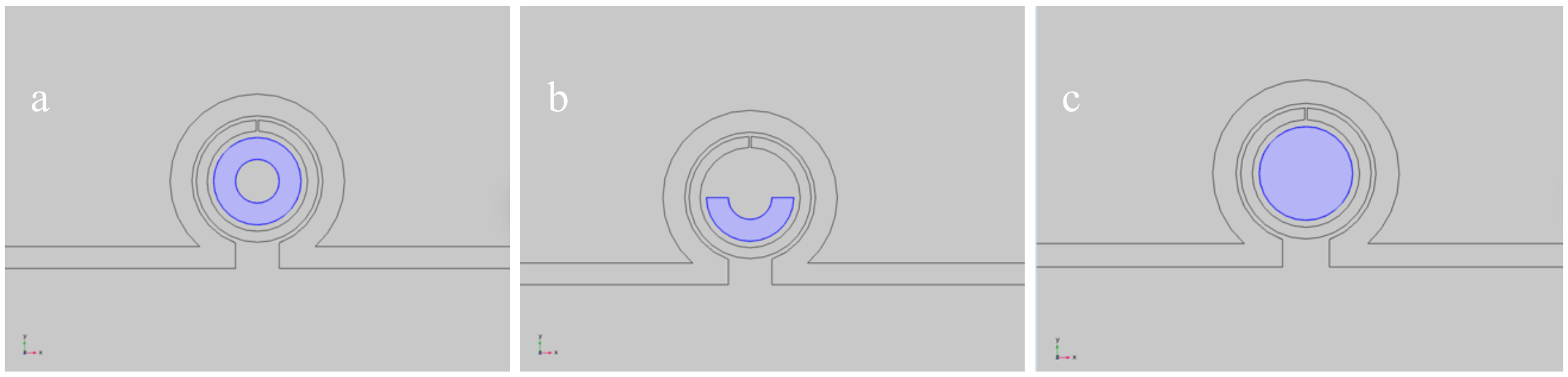

Figure 4: YIG geometries placed inside the optimized ASRR cavity: (a) full ring, (b) half ring, and (c) disk. Each configuration enables a different field-overlap condition affecting the coupling strength.

To quantify the coupling dynamics, the linewidths of both photon and magnon modes were extracted directly from the RF simulations. The photon linewidth was determined from the full width at half maximum (FWHM) of the uncoupled ASRR resonance, using the relation $\kappa_c = \frac{\omega_c}{2Q}$, yielding $\kappa_c = 28.8$ MHz. This value remains consistent across all geometries, as the cavity structure and resonance frequency were preserved during the optimization process. The FMR peak width in the simulated field-sweep spectra was used to calculate the magnon linewidth. The coupling strength $g$ was extracted from transmission spectra at the point of zero detuning from the FMR field, where the cavity and magnon uncoupled resonances coincide. The coupling strength $g$ is half of the frequency separation ($2g$) between two hybridized modes:

$$\omega_{\pm} = \frac{\omega_c+\omega_m}{2} \pm \sqrt{g^2 + \left(\frac{\Delta}{2}\right)^2} \qquad \Delta = \omega_c - \omega_m \ . \quad (13)$$

Once $g, \kappa_c, and\ \kappa_m$ were known, the cooperativity $C$ is calculated from Eq. (2). These parameters describe the coupling efficiency and allow consistent comparison among the different YIG geometries.

## 4.1 Full-Ring Geometry

The dependence of the transmission spectra on the bias magnetic field for the ASRR cavity integrated with the YIG full ring is shown in Fig. 5. Strong magnon–photon coupling is indicated by a distinct avoided crossing near 186 mT and a frequency of 5.48 GHz, as observed in the |S21| spectra in Fig. 6(a). The upper and lower magnon–photon polariton branches are split by approximately 230 MHz at zero detuning, corresponding to a coupling strength of $g\ =\ 115$ MHz. The magnon linewidth was determined from the uncoupled (detuned) region of the dispersion, This yields $\kappa_m/2\pi = 35$ MHz. These parameters result in cooperativity of $C \approx\ 13.1$, indicating that the system operates in the strong-coupling regime.

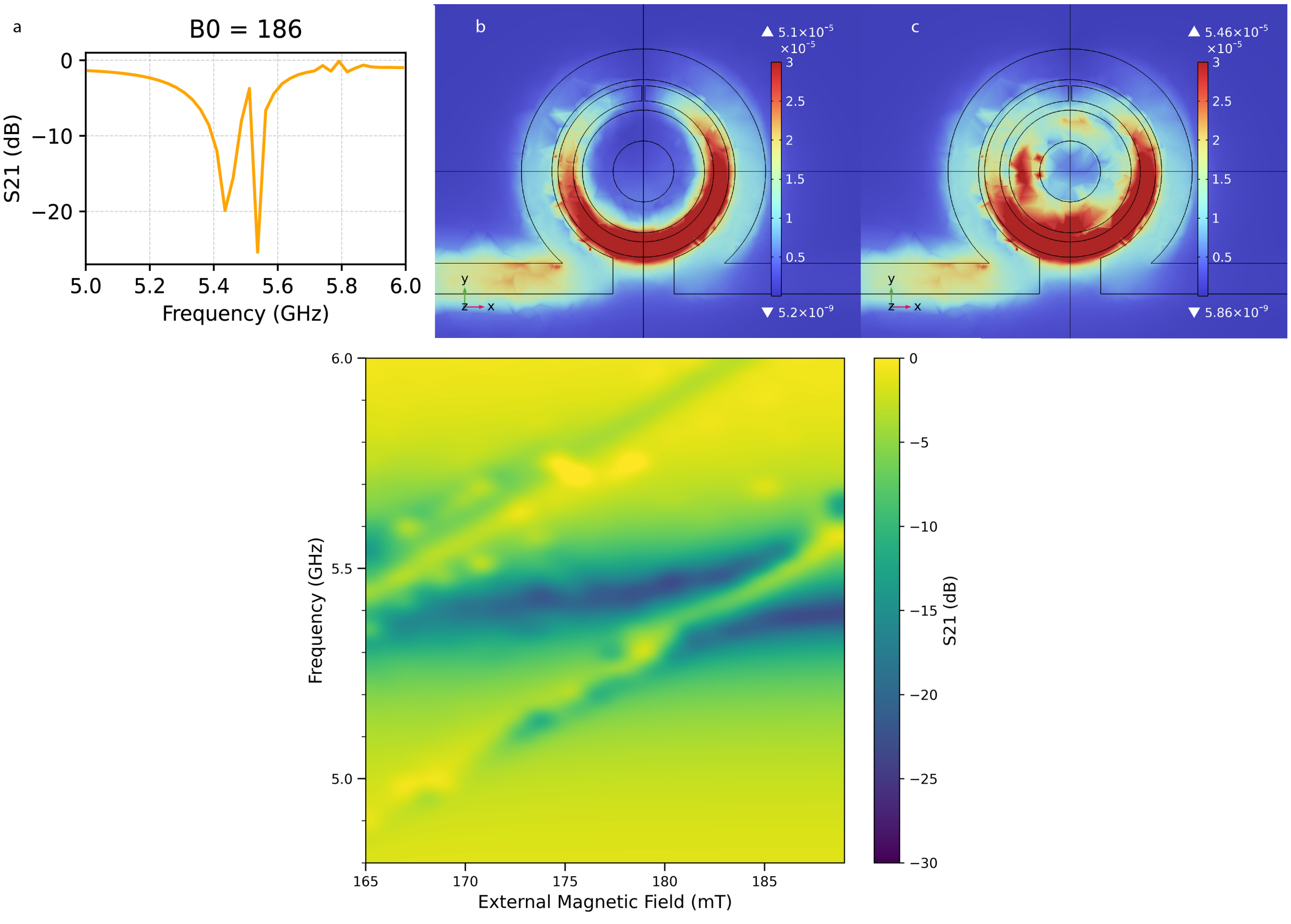

Figure 5: Magnon–photon coupling heatmap showing an avoided level crossing between the cavity microwave photon mode (at 5.48 GHz) and the uniform magnon mode in the YIG full-ring. The color scale represents the simulated transmission parameter $S_{21}$ (in dB). The anti-crossing occurs near $B_0$ =186 mT.

Figure 6: (a) Transmission spectrum |S21| of the ASRR cavity coupled to the YIG full ring at $B_0 = 186mT$, showing clear mode splitting due to magnon–photon hybridization. (b–c) Distribution of the magnetic field intensity of the upper and lower magnon-photon polariton branches at the coupling field.

### 4.2 Half-Ring Geometry

The half-ring geometry was then examined to investigate the impact of edge demagnetization and structural discontinuity. Removing a portion of the ring introduces open boundaries, which modify the uniformity of magnetization precession and the distribution of the demagnetizing field. The avoided crossing exhibits a frequency splitting of approximately 108 MHz at nearly the same bias field, i.e., 186 mT (see Fig. 7 and Fig. 8). At this field, the half-ring configuration exhibits two hybrid modes corresponding to the lower and upper branches at 5.41 GHz and 5.518 GHz, respectively (Fig. 7(a)). Despite a similar coupling strength, magnetic losses slightly increase, yielding $\kappa_m/2\pi = 30$ MHz. The resulting cooperativity remains comparable ($C \approx 13.5$), suggesting that increased magnetic damping and reduced mode uniformity slightly degrade the coupling efficiency, even though the microwave magnetic field intensity remains strong (Fig. 7(b)). Thus, compared to the full-ring configuration, the half-ring geometry indicates that open boundaries can introduce enhanced demagnetizing effects, which may lead to deviations from uniform precession, consistent with the modified field distributions observed in Fig. 7, however remaining similar cooperativity.

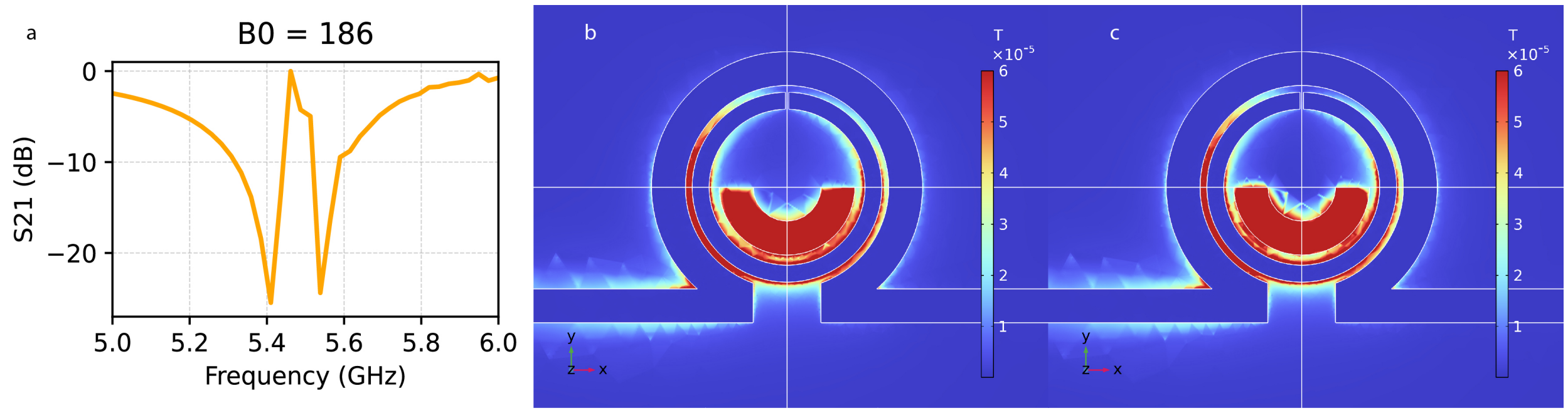

Figure 7: Transmission spectra of the ASRR–YIG half-ring configuration. (a) Avoided crossing observed at $B_0 = 186\ mT$. (b) Spatial distribution of the microwave magnetic field for the lower branch. (c) Spatial distribution for the upper branch.

To further analyze the nature of the hybrid modes, the magnetic flux density maps, taken in the mid-plane of the ASRR, reveal that the microwave magnetic field is mainly confined, particularly near the edges facing the feedline. In the upper branch, the microwave magnetic field shows enhanced localization within the YIG region, indicating a more magnon-like character. In contrast, for the lower branch, the field distribution exhibits a stronger presence in the ASRR structure, suggesting a more photon-like character. Although the contrast between the two modes is not strongly pronounced in the present visualization, this behavior remains consistent with the expected hybridization between magnon and photon modes.

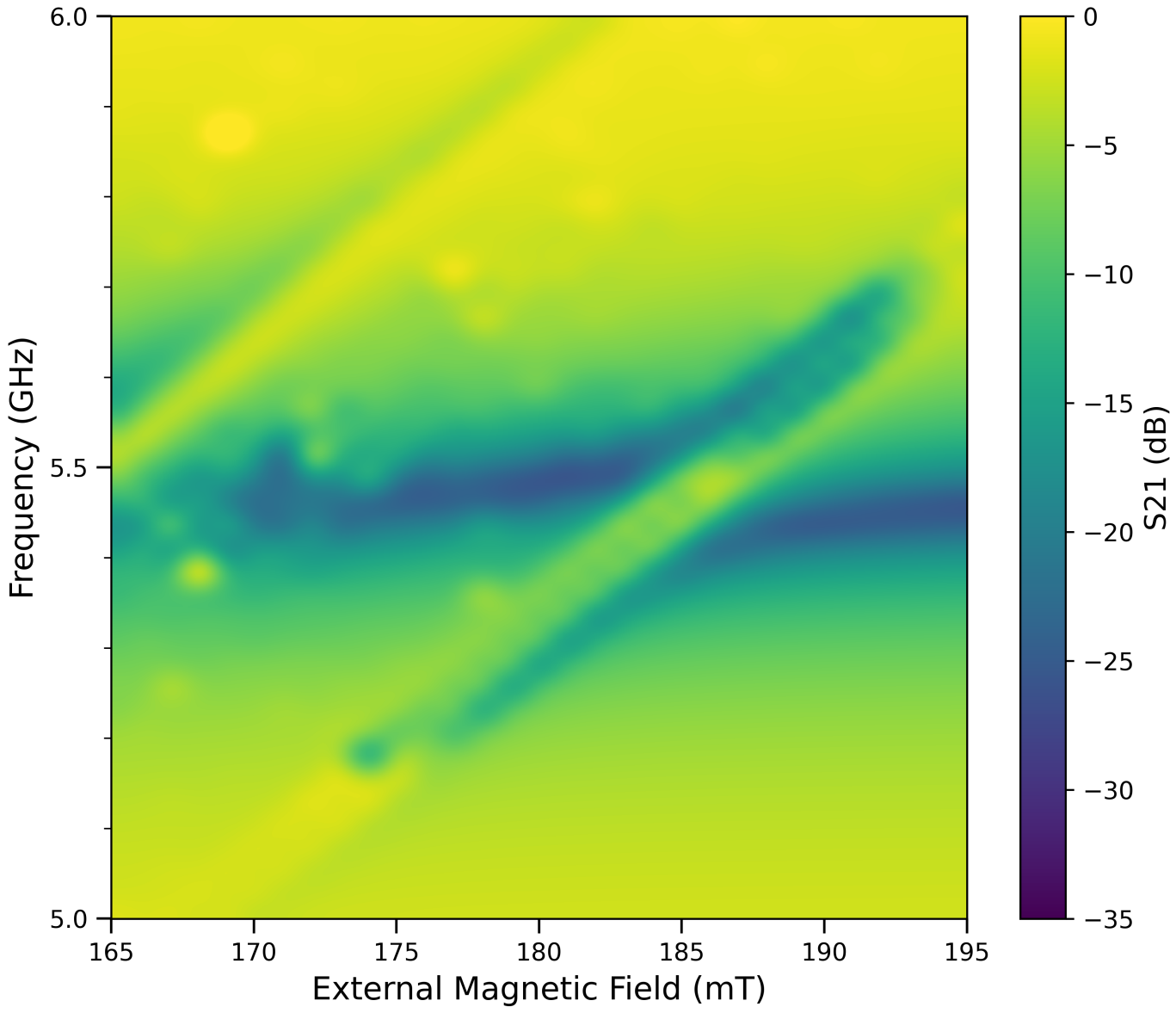

Figure 8: Heatmap of transmission as a function of bias magnetic field and frequency for the ASRR system with half-ring YIG. The avoided crossing near $B_0$ = 186 mT indicates strong magnon-photon coupling.

## 4.3 Disk Geometry

The disk geometry provides a compact configuration that maximizes spatial overlap with the ASRR's in-plane microwave magnetic field and corresponds to the largest magnetic volume. Its symmetric shape leads to a more uniform demagnetizing field distribution. In the disk geometry, the center of the cavity resonance frequency shifts to lower frequencies, see Figs. 9 and 10 from approximately 5.48 GHz to about 5.2 GHz. This shift is attributed primarily to the dielectric properties of the YIG disk, which has a relative permittivity $\varepsilon_{r,YIG} \approx$ 14–16, significantly higher than that of the RO4350B substrate ($\varepsilon_r = 3.48$) [10]. Placement of the disk in regions with non-negligible electric field increases the effective capacitance, lowering the resonance frequency, since it is proportional to $1/\sqrt{LC}$ [26]. While YIG's magnetic properties govern magnon–photon coupling via the microwave magnetic field, the effective permeability remains close to unity away from the ferromagnetic resonance condition, so its contribution to the cavity frequency shift is weak [10,30]. In this regime, dielectric contrast becomes the dominant mechanism. This is further confirmed by simulations in which the relative permittivity of YIG is artificially reduced to unity, demonstrating that dielectric loading by the YIG is the primary origin of the frequency shift.

According to the Kittel formula (Eq. 3), the ferromagnetic resonance occurs at approximately 132 mT at frequency of the microwave cavity, which agrees with the resonance obtained from the LLG analysis and the observed avoided crossing at 134 mT, with a mode splitting of $2g = 270$ MHz, ($g = 135$ MHz), as shown in Fig. 9 and Fig. 10. The extracted magnon decay rate is $\kappa_m/2\pi =$ 25 MHz, yielding the highest cooperativity ($C \approx 25.3$) among all geometries.

This strong enhancement stems primarily from three factors: (i) optimal field alignment, where the ASRR's in-plane microwave field efficiently overlaps with the disk's uniform Kittel mode; (ii) reduced edge demagnetization, as the absence of internal boundaries supports coherent precession; and (iii) the larger magnetic volume of the YIG disk. These features make the disk geometry attractive for compact, tunable magnon–photon hybrid devices, although this comes at the cost of a larger YIG volume.

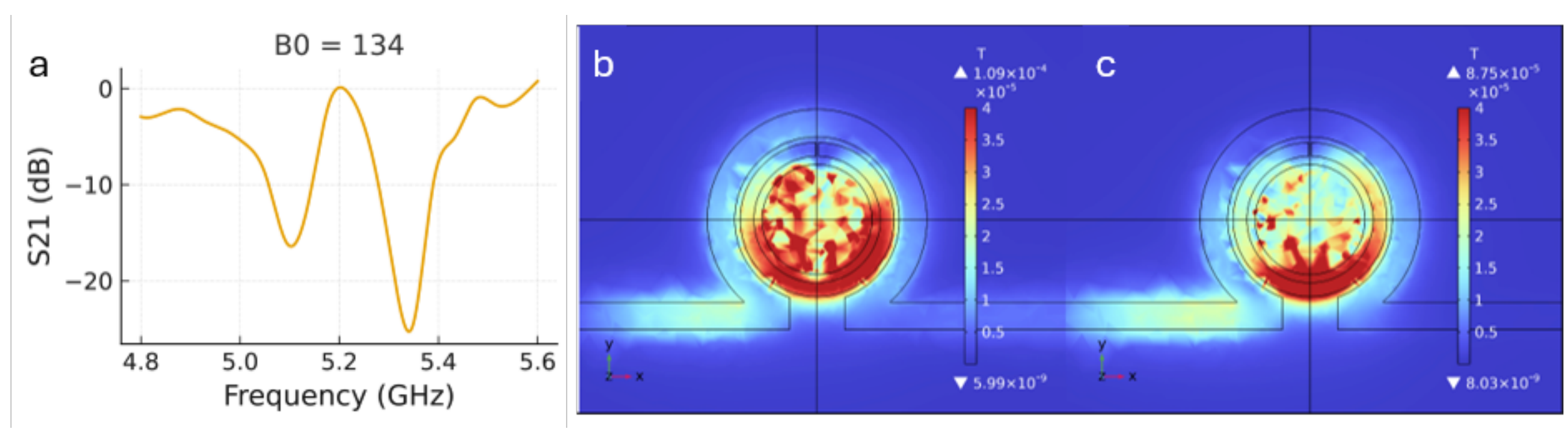

Figure 9: (a) Transmission spectra of the ASRR-YIG disk configuration at the avoided crossing occurs near $B_0 = 134\ mT$. (b) Magnetic field distribution of the lower branch and (c) magnetic field distribution of the upper branch at $B_0 = 134\ mT$

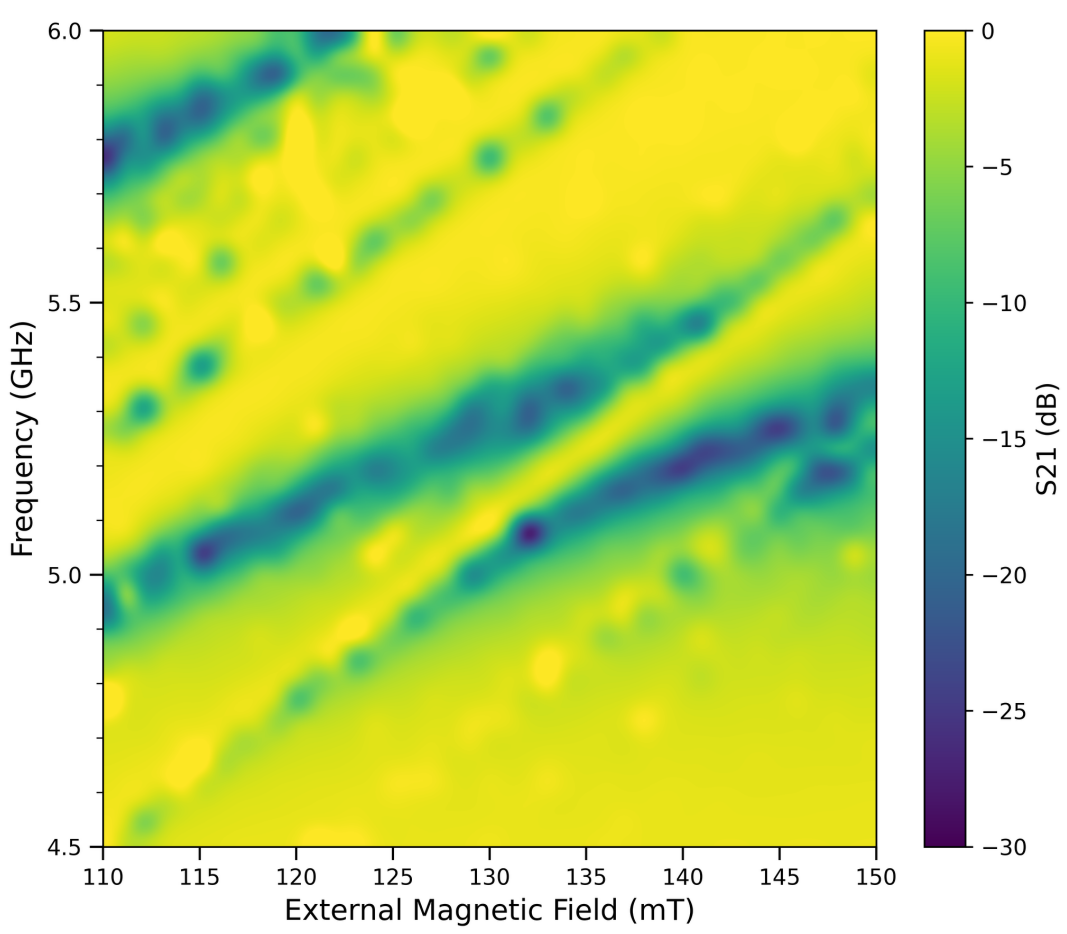

Figure 10: Heatmap of the transmission parameter $|S_{21}|$ as a function of the frequency and bias magnetic field for the ASRR coupled to a YIG disk. The avoided crossing near $B_0 = 134mT$ clearly indicates strong magnon-photon coupling.

Figure 9(b) shows the spatial distribution of the microwave magnetic field magnitude |B| in the YIG disk at a bias field of 134 mT (near maximum hybridization). The field exhibits strong, confinement throughout the disk volume, with noticeable enhancement near the ASRR gap and feedline. This behavior is consistent with excitation of the fundamental Kittel mode uniform, in-phase precession of magnetization around the in-plane bias field which corresponds to the dominant hybridized mode observed in the transmission spectra [3,9]. Minor secondary variations, which can arise from weak higher-order magnetostatic modes or edge effects, remain small compared to the main mode [10,29].

## 4.4 Comparative Summary

Table 1 summarizes the quantitative comparison between the three geometries. The results confirm that magnetic geometry plays an important role in determining both the coupling strength and cooperativity in planar ASRR–YIG systems.

| YIG Geometry | $B_0$ [mT] | **g** [MHz] | $\kappa_c$ [MHz] | $\kappa_m$ [MHz] | C | Qualitative Coupling Efficiency |
|---|---|---|---|---|---|---|
| Half Ring | 188 | 108 | 28 | 30 | 13.5 | Strong, slightly reduced due to edge demagnetization and small volume of YIG |
| Full Ring | 186 | 115 | 28 | 35 | 13.1 | Strong, well-balanced hybridization |
| Disk | 134 | 135 | 28 | 25 | 25.3 | Very strong, enhanced by field overlap and in-plane symmetry |

Table 1: The half-ring geometry exhibits strong coupling with reduced volume of the ferromagnet as compared to other cases, with similar cooperativity to the full-ring configuration. The disk geometry achieves the strongest coupling, owing to optimal field overlaps and its highest volume.

In the half-ring geometry, open edges modify the local demagnetizing field and slightly reduce the uniformity of magnetization precession. As a result, a small decrease in coupling strength is observed (from 115 MHz to 108 MHz), while the cooperativity remains comparable ($C \approx 13.5$) in the strong-coupling regime. In contrast, the disk geometry provides enhanced spatial overlap between the uniform in-plane magnetization and the localized microwave magnetic field. Its symmetric shape and high volume support coherent precession, leading to the largest Rabi splitting and strongest hybridization among the considered geometries.

To further assess intrinsic coupling efficiency, the collective coupling strength g was normalized using $g = g_0\sqrt{N}$, yielding the single-spin coupling strength $g_0 = g/\sqrt{N}$ [25]. The number of spins was estimated as $N = n_s V$, where $n_s \approx 4 \times 10^{27}$ m$^{-3}$ [10-12]. The results (Table 2) show that the half-ring exhibits a highest normalized coupling, suggesting optimized microwave and magnon field concentrations.

| Geometry | Volume V($m^3$) | Spins $N = n_s V$ | Coupling g (MHz) | Normalized coupling $g = g_0/\sqrt{N}$ |
|---|---|---|---|---|
| Full ring | $4.712 \times 10^{-9}$ | $1.885 \times 10^{19}$ | 115.00 | $2.65 \times 10^{-8}$ |
| Half ring | $2.356 \times 10^{-9}$ | $9.424 \times 10^{18}$ | 108.00 | $3.52 \times 10^{-8}$ |
| Disk | $6.283 \times 10^{-9}$ | $2.513 \times 10^{19}$ | 135.00 | $2.69 \times 10^{-8}$ |

Table 2: Comparison of volumes, spin numbers *N*, coupling strengths, and normalized per-spin coupling for the full ring, half ring, and disk YIG geometries.

The normalized coupling analysis shows that the full-ring geometry exhibits the lowest single-spin coupling efficiency, which can be attributed to the larger internal void and reduced concentration of the microwave magnetic field overlapping with the magnon mode. In contrast, the half-ring geometry achieves a higher normalized coupling $g_0$, despite having fewer spins, indicating enhanced local field concentration. However, this improvement is accompanied by increased magnetic damping as compared to disk geometry. The disk geometry, while not exhibiting the highest normalized coupling, achieves the largest collective coupling strength *g* due to its optimal spatial alignment with the cavity magnetic field and its largest magnetic volume.

# 5. Conclusion

In this work, a planar attached split-ring resonator (ASRR) cavity was designed and optimized to achieve strong magnon–photon coupling with YIG in a compact, on-chip platform. The optimized ASRR exhibits a resonance at 5.48 GHz (reduced to 5.2 GHz when the YIG disk is included) with enhanced magnetic-field confinement, reduced radiative losses, and an improved quality factor ($Q \approx 190$), making it well suited for hybrid spin–photon systems.

By integrating the ASRR with YIG elements of different geometries (full ring, half ring, and disk), we systematically investigated the role of magnetic shape in determining coupling strength and dissipation. The full-ring geometry provides balanced performance, with a coupling strength of 115 MHz and cooperativity $C \approx 13.1$. The half-ring maintains strong coupling ($g$ = 108 MHz, $C \approx 13.5$) but similar as a full ring, exhibits increased magnetic losses due to edge-induced demagnetization. In contrast, the disk geometry achieves the strongest interaction ($g$ = 135 MHz, $C \approx 25.3$) and operates at lower bias fields. However, the strongest coupling per spin is obtained for the YIG geometry of the smallest volume, i.e., half ring, when placed inside the ASRR at the edge close to the feeding line. The results demonstrate that magnon–photon coupling in planar systems is governed not only by magnetic volume, but critically by the spatial and directional matching between the microwave magnetic field and the magnetization dynamics. While reduced volumes can enhance local coupling efficiency, maximizing collective coupling requires strong fields overlap. These findings establish geometric engineering as a key design principle for optimizing magnon–photon interactions in planar platforms, providing a pathway toward scalable, lithography-compatible hybrid magnonic and photonic devices.

# Acknowledgements

The research has received financial support from the National Science Center of Poland, OPUS-LAP grant no. 2020/39/I/ST3/02413.